\documentclass[floatfix,twocolumn,aps,prd,showpacs,10pt,superscriptaddress]{revtex4-1}
\usepackage{amsfonts}
\usepackage{epsfig}
\usepackage{graphicx}
\usepackage[latin1]{inputenc}
\usepackage{float}
\usepackage{color}
\usepackage{amsmath,amssymb}
\usepackage{natbib}
\usepackage{hyperref}
\hypersetup{colorlinks=true, urlcolor=blue,linkcolor=blue, citecolor=blue}
\begin{document}
\title{Quantum forces in cavities}
\author{E. Cavalcanti}
\email{erichcavalcanti@gmail.com}
\author{A.P.C. Malbouisson}
\email{adolfo@cbpf.br}
\affiliation{Centro Brasileiro de Pesquisas F\'{i}sicas/MCTI, 22290-180 Rio de Janeiro, Brazil.}
\begin{abstract}
We present an alternative route to investigate quantum forces in a cavity, for both a free scalar field and a free fermionic field. Con\-si\-de\-ring a generalized Matsubara procedure, we compute the quantum pressure on the boundaries of a compactified space in a thermal bath. Also, considering both periodic and antiperiodic boundary conditions in the cavity, we discuss the attractive or repulsive character of the quantum forces. Our results highlight the relevance and interdisciplinarity of the generalized Matsubara procedure as a theoretical platform to manage quantum phenomena in nontrivial topologies.
\end{abstract}
\maketitle
\section{Introduction} 
\label{intro} 

Quantum forces emerge as a macroscopic manifestation of the existence of quantum vacuum fluctuations. Perhaps, the so-called Casimir forces, originally described as the attraction of two conducting, neutral plates in a vacuum, induced by changes in the zero-point energy of the electromagnetic field \cite{Casimir-1948}, represent the most popularized among them (see Refs. \cite{Mostepanenko-Trunov-1997, Milton-2001, Bordag-etal-2009, CasimirPhysics-2011, Lamoreaux-PRL-1997, Mohideen-Roy-PRL-1998, Chan-etal-Science-2001, Decca-etal-PRL-2003, Intravaia-etal-Science-2001, Woods-etal-App-Sci-2021} for more details on theoretical and experimental aspects of the Casimir Physics).

The understanding of the sign of the Casimir force, that is, whether it is attractive or repulsive plays a significant role in the specialized literature. In the initially studied models, the force was purely attractive, and this was suggested as one critical aspect of the stability of matter. However, later on, this was discarded because the force turns out to be repulsive in spherical geometry, as shown by Ref.~\cite{Boyer-PR-1968}. Over the last decades, many studies indicated that the sign of the Casimir force is related to the choice of boundary conditions rather than the physical model under investigation (for an extensive list of references see Refs.~\cite{Flachi-etal-PRL-2017,Gundersen-Ravndal-Ann-Phys-1988}).

In the past decades, one suggested path to investigate nontrivial topologies is to employ the methodology of quantum field theories on toroidal topologies \cite{Khanna-et-al-Thermal-Book-2009,Khanna-et-al-Phys-Rep-2014}. This method allows us to study both thermal and finite-size effects on a wide range of applications: spontaneous symmetry breaking \cite{APCM-et-al-NPB-2002},
second-order phase transitions in superconducting films, wires, and grains 
\cite{APCM-et-al-MPLA-2005,Abreu-et-al-JMP-2005,APCM-et-al-JMP-2009},
finite-size effects in the presence of magnetic fields, finite chemical
potential in first-order phase transitions \cite{Correa-et-al-PLA-2013}, the Casimir effect \cite{Ford-PRD-1980, Kleinert-Zhuk-TMP-1996, JCSilva-PRA-2002, Queiroz-et-al-Annals-of-Phys-2005, Queiroz-et-al-Annals-of-Phys-2006, Ahmadi-et-al-PRD-2002, Tomazelli-et-al-IJTP-2006, Bezerra-etal-PRD-2011-I, Bezerra-etal-PRD-2011-II, Bezerra-Mota-PRD-2014, Mota-Bezerra-PRD-2015, Farias-Mota-PLB-2020} and related phenomena \cite{Abrantes-et-al-PRD-2018, Santos-Khanna-EPL-2019, Santos-Khanna-Ann-Phys-2021} and other quantum effects \cite{Cavalcanti-etal-PRD-2019, Mogliacci-et-al-PRD-2020, Cavalcanti-PRD-2021}.

From finite-temperature field theory, it is well-known that one
way to introduce thermal effects is to consider the imaginary-time Matsubara formalism. As the name suggests, we impose that the dimension related to imaginary time (Wick-rotated variable, so we are in Euclidean `space-time' instead of Minkowski space-time) has a finite extension $\beta = 1/T$, with a periodic (antiperiodic) boundary condition for bosons (fermions) as imposed by the KMS condition~\cite{Khanna-et-al-Thermal-Book-2009}. This procedure was later on extended by Birrell and Ford \cite{Birrel-Ford-PRD-1980} to describe field theories in spaces with finite geometries and has been generalized to what came to be known as quantum field theories on toroidal topologies \cite
{APCM-et-al-NPB-2002, Khanna-et-al-Annals-of-Physics-2009,
Khanna-et-al-Thermal-Book-2009, Khanna-et-al-Annals-of-Physics-2011,
Khanna-et-al-Phys-Rep-2014,Cavalcanti-PRD-2021}. This procedure can also be called a \textit{generalized Matsubara formalism}. 
In general, this technique consists in considering quantum field defined on $D$ Euclidean dimensions with $d$ of them ($1\le d\le D$) \textit{compactified} with a characteristic length. The compactification is introduced through the imposition of boundary conditions on the components of the fields. One of these dimensions is compactified in a circumference of length $\beta=1/T$ and the boundary conditions are restricted by the KMS condition to be periodic or antiperiodic, depending on the statistics of the field. The compactified spatial dimensions are restricted to a characteristic length $L_i$ ($i=1,\ldots,d-1$) and can be interpreted as boundaries of the Euclidean space. In the special case where we impose periodic boundary conditions the topology of the space is of the type $\left( \mathbb{S}^{1}\right) ^{d}\times \mathbb{R}^{D-d}$ \cite
{Khanna-et-al-Thermal-Book-2009, Khanna-et-al-Phys-Rep-2014}.

Here, we propose an alternative method to obtain quantum forces in nontrivial topologies, by employing the generalized Matsubara formalism. We start with the so-called \textquotedblleft local formulation\textquotedblright , introduced in \cite {Brown-Maclay-Phys-Rev-1969}, in which the pressure is associated with the (3,3)-component of the energy-momentum tensor. Then we employ a general formalism of field theories on toroidal spaces as in Ref.~\cite{Khanna-et-al-Phys-Rep-2014}. This methodology, valid for several simultaneously compactified dimensions follows the method originally used by Elizalde and
Romeo \cite{Elizalde-Romeo-JMP-1989, Elizalde-Romeo-JMP-1990} for the computation of the Casimir energy. The generalization to an arbitrary number of compactified dimensions is useful, for instance, for thermal field theories with a finite spatial
extension. For example, a unified approach for Casimir cavities in a thermal bath requires that both imaginary-time dimension and a spatial one are compactified.

We stress that during this paper our computation with the toroidal formalism implements by construction periodic and/or antiperiodic boundary conditions. For the imaginary time (circumference of
length $\beta=1/T$) we use periodic boundary conditions for bosons and antiperiodic boundary conditions for fermions. While for the third spatial coordinate (circumference of length $L$%
) we can employ both. Moreover, we can obtain results from other boundary conditions using the information from periodic ones \cite{Khanna-et-al-Phys-Rep-2014,Cavalcanti-etal-PRD-2019b}. For instance, the pressure for Dirichlet boundary conditions (much studied in the literature) for parallel planes at distance $a$ (Ref.~\cite{Brown-Maclay-Phys-Rev-1969}) is recovered from the pressure for periodic boundary conditions for parallel planes at a distance $L=2a$.

The paper is organized as follows. In the next section, we link the quantum pressure to the vacuum expectation value of the energy-momentum tensor for both a scalar and fermionic field in $D$ dimensions of the Euclidean space. 
In subsection \ref{EMT_for_scalar_fields} we use the point-splitting technique and rewrite the energy-momentum tensor in terms of the free scalar propagator in Fourier space. 
In subsection \ref{VacuumPressureInToroidalSpace} we obtain a corresponding expression for the quantum pressure in a cavity (when one of the spatial dimensions is compactified with a finite
extension). The computation of the quantum vacuum pressure follows a path similar to
that of the Elizalde--Romeo method \cite{Elizalde-Romeo-JMP-1989}, leading
to a well-known result from the literature. 
In section \ref{ThEf} we compute the Casimir pressure in a cavity in the presence of a thermal bath. This can also be compared with results found in the literature obtained from other techniques. We also exhibit, see Fig.~\ref{fig:scalarphasediagram}, the cross-over between an attractive and a repulsive scenario. In section \ref{ThEf_Fermion} we repeat the computation for a fermionic model and also obtain a cross-over between an attractive and repulsive pressure, see Fig.~\ref{fig:fermionphasediagram}.
In section \ref{finalremark} we present our final comments. Throughout this
paper, we consider $\hbar =c=k_{B}=1$.

\section{A scalar field model}
\label{sec_the_model}
\subsection{Energy-momentum tensor for scalar fields} 
\label{EMT_for_scalar_fields} 

We start by writing the Euclidean Lagrangian for both a free scalar field and a free fermionic field in a $D$-dimensional space, 
\begin{subequations}
\begin{equation}
\mathcal{L}_{E}^{\text{scalar}}=\frac{1}{2}\left( \partial _{\mu }\phi \right) ^{2}+
\frac{1}{2}m^{2}\phi ^{2},  
\label{eq:lagrangian}
\end{equation}
\begin{equation}
\mathcal{L}_{E}^{\text{fermion}}=\partial_\mu \bar \psi \partial^\mu \psi+
m\bar\psi \psi,  
\label{eq:lagrangian_fermion}
\end{equation}
\end{subequations}
\noindent where $m$ is the mass of the quanta of the field. To simplify notation we use the same mass for both the scalar and the fermionic fields. With the help of the point-splitting technique, the vacuum expectation value of the canonical energy-momentum tensor 
$T_{\mu \nu }$ can be written as \cite{Khanna-et-al-Phys-Rep-2014}, 
\begin{subequations}
\begin{eqnarray}
\mathcal{T}_{\mu \nu }^{\text{scalar}} &=&\left\langle 0\left\vert T_{\mu \nu }\right\vert
0\right\rangle  \nonumber \\
&=&\lim_{x^{\prime }\rightarrow x}\!\mathcal{O}_{\mu \nu }^{\text{scalar}}\left( x,x^{\prime
}\right) \!\left\langle 0\left\vert {T}\left\{\phi \left( x\right) \phi \left(
x^{\prime }\right)\right\} \right\vert 0\right\rangle \!,
\label{eq:energymomentumtensor}
\end{eqnarray}
\begin{multline}
\mathcal{T}_{\mu \nu }^{\text{fermion}} =\\{\text{tr}}\Bigg\{\lim_{x^{\prime }\rightarrow x}\!\mathcal{O}_{\mu \nu }^{\text{fermion}}\left( x,x^{\prime
}\right) \!\left\langle 0\left\vert {T}\left\{\psi \left( x\right) \bar\psi \left(
x^{\prime }\right)\right\} \right\vert 0\right\rangle \!
\Bigg\},
\label{eq:energymomentumtensor_fermion}
\end{multline}
\end{subequations}
where ${T}$ denotes the time-ordered product of field operators, $\text{tr}$ is a trace over the internal indices of the spinor, and $\mathcal{O}_{\mu \nu }\left( x,x^{\prime }\right) $ is a differential operator given by \cite{Khanna-et-al-Phys-Rep-2014},
\begin{subequations}
\begin{equation}
\mathcal{O}_{\mu \nu }^{\text{scalar}}\left( x,x^{\prime }\right) =\partial _{\mu }\partial
_{\nu }^{\prime }-\frac{1}{2}\delta _{\mu \nu }\left(\partial _{\sigma
}\partial _{\sigma }^{\prime }+m^{2}\right) , 
\label{eq:smart_operator}
\end{equation}
\begin{equation}
\mathcal{O}_{\mu \nu }^{\text{fermion}}\left( x,x^{\prime }\right) = \frac{1}{4} \left( 
\gamma_\mu^E \left(\partial_\nu^\prime - \partial_\nu\right)
+ \gamma_\nu^E \left(\partial_\mu^\prime - \partial_\mu\right)
\right) , 
\label{eq:smart_operator_fermion}
\end{equation}
\end{subequations}

\noindent where $\partial _{\mu }$ and $\partial _{\mu }^{\prime }$ are derivatives
acting on $x^{\mu }$ and $x^{\prime \mu }$, respectively, $\delta _{\mu\nu }$ represents the components of the metric tensor of the Euclidean space (Kronecker delta) and $\gamma_\mu^E$ are the Euclidean Dirac matrices~\cite{Zinn-Justin-Book-2002}. Defining the Euclidean Green function of the scalar field as
$G\left( x-x^{\prime }\right) = \left\langle 0\left\vert 
{T}\left\{ \phi\left( x\right) \phi \left( x^{\prime }\right) \right\} 
\right\vert 0\right\rangle $, we obtain 
\begin{equation}
\mathcal{T}_{\mu \nu }^{\text{scalar}}=\lim_{x^{\prime }\rightarrow x}\mathcal{O}_{\mu \nu
}^{\text{scalar}}\left( x,x^{\prime }\right) G\left( x-x^{\prime }\right) .
\label{eq:energy_momentum_tensor_Euclidean_Green_function}
\end{equation}
Considering the Fourier integral of the Euclidean Green function in momentum
space, 
\begin{equation}
G\left( x-x^{\prime }\right) =\int_{-\infty }^{\infty }\frac{d^{D}k}{\left(
2\pi \right) ^{D}}\frac{1}{k^{2}+m^{2}}e^{ik\cdot \!\left( x-x^{\prime
}\right) },  
\label{eq:fourierGreenfunction}
\end{equation}
where $k$ and $x$ are $D$-dimensional vectors, we are able to rewrite the
vacuum expectation value of the energy-momentum tensor in Eq.~(\ref%
{eq:energy_momentum_tensor_Euclidean_Green_function}) as 
\begin{equation}
\mathcal{T}{}_{\mu \nu }^{\text{scalar}}=\int_{-\infty }^{\infty }\frac{d^{D}k}{\left( 2\pi
\right) ^{D}}\left[ \frac{k_{\mu }k_{\nu }}{k^{2}+m^{2}}-\frac{1}{2}\delta
_{\mu \nu }\right] .  
\label{eq:standard_tensor}
\end{equation}
In the fermionic scenario we define the Euclidean Green function as $S\left( x-x^{\prime }\right) = \left\langle 0\left\vert 
{T}\left\{ \psi\left( x\right) \bar\psi \left( x^{\prime }\right) \right\} 
\right\vert 0\right\rangle $. Which can be written in terms of the Green function for the scalar field,
\begin{equation}
S\left( x-x^{\prime }\right) = (\gamma^E_\sigma\partial^\sigma+m)G\left( x-x^{\prime }\right),  
\label{eq:fourierGreenfunction_fermion}
\end{equation}
\noindent therefore we obtain that the vacuum expectation value for the fermionic energy-momentum tensor is 
\begin{equation}
\mathcal{T}_{\mu \nu }^{\text{fermion}}=\int_{-\infty }^{\infty }\frac{d^{D}k}{\left( 2\pi
	\right) ^{D}}\left[ \frac{Nk_{\mu }k_{\nu }}{k^{2}+m^{2}}\right] .  
\label{eq:standard_tensor_fermion}
\end{equation}
\noindent Here we used that $\text{tr}\left(\gamma_\mu^E \gamma_\nu^E\right) = (1/2) \text{tr} \left\{\gamma_\mu^E, \gamma_\nu^E\right\} = (1/2) \delta_{\mu\nu} \text{tr} \mathbb{I} = N \delta_{\mu\nu}$, where $N = 2^{\lfloor D/2 \rfloor}$ is the size of the gamma matrix.

It is sufficient to consider the 33-component of the energy-momentum tensor to obtain the quantum force per unit area resulting from a topological constraint imposed by periodic boundary conditions on the field at the parallel plates (taken as infinite planes)
separated by a fixed distance $L$ in the $x_{3}$-direction.
From Eq.~(\ref{eq:standard_tensor}) and Eq.~\eqref{eq:standard_tensor_fermion}, it is straightforward to write the bulk
expression 
\begin{subequations}
	\begin{equation}
	\mathcal{T}_{33}^{\text{scalar}}=\frac{1}{2}\int_{-\infty }^{\infty }\frac{d^{D}k}{\left(
		2\pi \right) ^{D}}\left[ \frac{k_{3}^{2}-\left( k_{\bot }^{2}+m^{2}\right) }{%
		k_{3}^{2}+k_{\bot }^{2}+m^{2}}\right] ,
	\label{eq:previous_pressure_expression}
	\end{equation}
	\begin{equation}
	\mathcal{T}_{33}^{\text{fermion}}=2^{\lfloor D/2 \rfloor}\int_{-\infty }^{\infty }\frac{d^{D}k}{\left(
		2\pi \right) ^{D}}\left[ \frac{k_{3}^{2}}{%
		k_{3}^{2}+k_{\bot }^{2}+m^{2}}\right] ,
	\label{eq:previous_pressure_expression_fermion}
	\end{equation}
\end{subequations}
where $k^{2}=k_{3}^{2}+k_{\bot }^{2}$, and $k_{\bot }$ refers to the ($D-1$%
)-dimensional vector component orthogonal to the 3-direction in Fourier space.
\subsection{Quantum vacuum pressure in toroidal space - scalar field}
\label{VacuumPressureInToroidalSpace} 

In this section, we investigate the Casimir pressure for the particular case
of just one compactified spatial dimension ($d=1$). We restrict ourselves to the scalar scenario and discuss the fermionic model in Sec.~\ref{ThEf_Fermion}
Let us call $\mathcal{T}_{33,\mathit{c}}$ the response of vacuum
fluctuations on the plates, viewed as a topological constraint.
We perform this using the compactification of just one spatial
dimension. To obtain the Casimir pressure that acts on the boundary
of the compactified space, we shall use the generalized Matsubara procedure,
which is the original contribution of the present manuscript. Basically, in
the general case, the technique consists of the replacement of integrals in
momentum space by sums, namely, 
\begin{equation*}
\int \frac{dk_{j}}{2\pi }\rightarrow \frac{1}{L_{j}}\sum_{n_{j}=-\infty
}^{+\infty }
\end{equation*}
where the index $j$ assumes the values $j=1,2,\ldots ,D-1$, the momentum
coordinate $k_{j}$ exhibits discrete values, 
\begin{equation*}
k_{j}=k_{n_{j}}=\frac{2\pi n_{j}}{L_{j}} + \frac{\theta_j \pi}{L_j},
\end{equation*}
and $L_{j}$ refer to the finite extension of each of the $j$ spatial
dimensions (compactification of $D-1$ spatial coordinates). The parameter $\theta_j$ allows us to consider both periodic ($\theta_j = 0$) or antiperiodic ($\theta_j=1$) boundary conditions in space. For practical
purposes, let us compactify just the $x_{3}$-component of the vector $x$.
With these ideas in mind, the generalized Matsubara formalism enables us to
substitute the bulk expression of Eq.~(\ref{eq:previous_pressure_expression}) with the following one: 
\begin{equation}
\mathcal{T}_{33,\mathit{c}}^{\text{scalar}}=\!\frac{1}{2L}\!\sum_{n=-\infty }^{+\infty
}\!\int_{-\infty }^{\infty }\frac{d^{D-1}k_{\bot }}{\left( 2\pi \right)
^{D-1}}\!\left[ \frac{k_{n}^{2}-\left( k_{\bot }^{2}+m^{2}\right) }{%
k_{n}^{2}+k_{\bot }^{2}+m^{2}}\right] \!.\;\;\;
\label{eq:tensor_matsubarizado_espacialmente}
\end{equation}
Using  well-known results from dimensional regularization techniques, we have  
\begin{eqnarray}
\int_{-\infty }^{\infty }\frac{d^{D}k}{\left( 2\pi \right) ^{D}}\frac{1}{%
\left[ k^{2}+b^{2}\right] ^{s}} &=&\frac{1}{\left( 4\pi \right) ^{\frac{D}{2}%
}}\frac{\Gamma \left( s-\frac{D}{2}\right) }{\Gamma \left( s\right) }  
\nonumber
\\
&&\times 
\left( \frac{1}{b^{2}}\right) ^{s-\frac{D}{2}},
\label{eq:firstmultidimensionalintegral}
\end{eqnarray}%
\begin{eqnarray}
\int_{-\infty }^{\infty }\frac{d^{D}k}{\left( 2\pi \right) ^{D}}\frac{k^{2}}{%
\left[ k^{2}+b^{2}\right] ^{s}} &=&\frac{D}{2}\frac{1}{\left( 4\pi \right) ^{%
\frac{D}{2}}}\frac{\Gamma \left( s-\frac{D}{2}-1\right) }{\Gamma \left(
s\right) }  \nonumber \\
&&\times \left( \frac{1}{b^{2}}\right) ^{s-\frac{D}{2}-1},
\label{eq:secondmultidimensionalintegral}
\end{eqnarray}%
so that we obtain 
\begin{eqnarray}
\mathcal{T}_{33,\mathit{c}}^{\text{scalar}} &=&\left\{ f_{s}\left( \nu ,L\right)
\sum_{n=-\infty }^{+\infty }\left[ \frac{\left( a\left(n+\theta/2\right)^{2}-c^{2}\right) \Gamma
\left( \nu \right) }{\left( a\left(n+\theta/2\right)^{2}+c^{2}\right) ^{\nu }}\right. \right. 
\nonumber  \label{eq:tensor_pre_summation} \\
&&\left. -\left. \frac{\left( s-\nu \right) \Gamma \left( \nu -1\right)}{%
\left( a\left(n+\theta/2\right)^{2}+c^{2}\right) ^{\nu -1}}\right] \right\} _{\!\!s=1}\!\!,  
\end{eqnarray}%
where $a=L^{-2}$, $c=m/{2\pi }$, $\nu =s-\left( D-1\right) /2$, and $%
f_{s}\left( \nu ,L\right) $ a function given by 
\begin{equation}
f_{s}\left( \nu ,L\right) =\frac{1}{2L}\frac{1}{\left( 4\pi \right) ^{s-\nu
}\left( 2\pi \right) ^{2\left( \nu -1\right) }\Gamma \left( s\right) }.
\label{f_vacuum}
\end{equation}%
Adding and subtracting the term $c^{2}\Gamma \left( \nu \right) $ to the
numerator of the first term on the right-hand side of Eq.~(\ref%
{eq:tensor_pre_summation}), we obtain 
\begin{eqnarray}
\mathcal{T}_{33,\mathit{c}}^{\text{scalar}} &=&\Bigg\{ f_{s}\left( \nu ,L\right)  \Gamma\left(\nu-1\right)
 \Bigg[
\sum_{n=-\infty }^{+\infty }\frac{\left( 2\nu -s-1\right)}{\left(
a\left(n+\theta/2\right)^{2}+c^{2}\right) ^{\nu -1}}  \nonumber
\label{eq:tensor_more_than_pre_summation} \\
&&\left. -\left.  \sum_{n=-\infty }^{+\infty }%
\frac{2c^{2}\left( \nu -1\right)}{\left( a\left(n+\theta/2\right)^{2}+c^{2}\right) ^{\nu }}\right] \right\} _{\!\!s=1}\!\!,
\end{eqnarray}%
where we have used the relation $\Gamma \left( \nu \right) =\left( \nu -1\right)
\Gamma \left( \nu -1\right) $. Recalling the general definition of the
multidimensional Epstein--Hurwitz zeta function \cite%
{Elizalde-Romeo-JMP-1989, Kirsten-JMP-1994, Elizalde-etalBook-1994,
Elizalde-etalBook-1995}, 
\begin{eqnarray}
Z_{d}^{c^{2}}\!\left( \nu \;;\mathbf{a}\;;\boldsymbol{\theta}\right)
&=&\!\!\!\sum_{n_{1},\ldots ,n_{d}=-\infty }^{+\infty }\!\!\!\left(
c^{2} + \sum_{i=1}^d a_i (n_i + \theta_i/2)^2\right) ^{-\nu },  \nonumber
\label{eq:multidimensional_Epstein_Hurwitz_zeta_function} \\
&&
\end{eqnarray}%
In the particular case of one-dimensional compactification ($d=1$), the above formula
simplifies to 
\begin{equation}
Z_{1}^{c^{2}}\!\left( \nu \;;a\;;\theta\right) =\sum_{n=-\infty }^{+\infty }\left(
a\left(n+\theta/2\right)^{2}+c^{2}\right) ^{-\nu }.
\label{eq:onedimensional_Epstein_Hurwitz_zeta_function}
\end{equation}%
Substituting the previous expression into Eq.~(\ref%
{eq:tensor_more_than_pre_summation}), the pressure can then be rewritten as 
\begin{eqnarray}
\mathcal{T}_{33,\mathit{c}}^{\text{scalar}} &=&\left\{ f_{s}\left( \nu ,L\right) \Gamma\left(\nu-1\right) \left[
\left( 2\nu -s-1\right) Z_{1}^{c^{2}}\!\left( \nu -1;a;\theta\right) \right. \right.
\nonumber \\
&&\left. -\left. 2c^{2}\left( \nu -1\right) Z_{1}^{c^{2}}\!\left( \nu
;a;\theta\right) \right] \right\} _{\!\!s=1}.
\label{eq:tensor_more_than_pre_summation2}
\end{eqnarray}%
Following Ref.~\cite{APCM-et-al-NPB-2002}, these zeta functions can be
evaluated on the whole complex plane by means of an analytic continuation along lines similar to those described in Refs. \cite{Elizalde-Romeo-JMP-1989, Elizalde-Romeo-JMP-1990, Kirsten-JMP-1994, Elizalde-etalBook-1994, Elizalde-etalBook-1995} leading to   
\begin{eqnarray}
Z_{d}^{c^{2}}\left( \nu ;\mathbf{a}; \boldsymbol{\theta}\right) &=&\frac{2\pi ^{\frac{d}{%
2}}}{\sqrt{a_{1}\cdots a_{d}}\;\Gamma \left( \nu \right) }\left[ \frac{1}{%
2c^{2\nu -d}}\Gamma \left( \nu -\frac{d}{2}\right) \right.  \nonumber \\
&&+2\sum_{j=1}^{d}\sum_{n_{j}=1}^{\infty }\left( \frac{\pi n_{j}}{c\sqrt{%
a_{j}}}\right) ^{\nu -\frac{d}{2}} \cos(\theta_j \pi n_j)  \nonumber \\
&&\times K_{\nu -\frac{d}{2}}\left( 2\pi c\frac{n_{j}}{\sqrt{a_{j}}}\right)
+\cdots  \nonumber \\
&&+2^{d}\sum_{n_{1},\ldots ,n_{d}=1}^{\infty }\left( \frac{\pi }{c}\sqrt{%
\sum_{i=1}^d\frac{n_{i}^{2}}{a_{i}}}\right) ^{\nu -\frac{%
d}{2}}  \nonumber \\
&&\times \left(\prod_{i=1}^{d} \cos(\theta_i\pi n_i)\right) K_{\nu -\frac{d}{2}}\left( 2\pi c\sqrt{\sum_{i=1}^d\frac{n_{i}^{2}}{a_{i}}}
\right),  \nonumber \\
&&  \label{eq:analytical_continuation_of_MultiZeta}
\end{eqnarray}%
where $K_{\nu }\left( z\right) $ denotes modified Bessel functions of the
second kind. For $d=1$, the analytical continuation above is reduced to 
\begin{multline}
Z_{1}^{c^{2}}\!\left( \nu ;a;\theta\right) \frac{2\pi ^{\frac{1}{2}}}{\sqrt{a}%
\;\Gamma \left( \nu \right) }\left[ \frac{1}{2c^{2\nu -1}}\Gamma \left( \nu -%
\frac{1}{2}\right) \right.  \\
\left. +2\sum_{n=1}^{\infty }\left( \frac{\pi n}{c\sqrt{a}}\right) ^{\nu -%
\frac{1}{2}}\!\cos(\theta\pi n)K_{\nu -\frac{1}{2}}\left( 2\pi c\frac{n}{\sqrt{a}}\right) %
\right] .  
\end{multline}%
After some algebraic manipulations, we notice the presence of terms which
are independent of the variable $L$, and for this reason are considered
unphysical. Neglecting these terms, we can show that 
\begin{multline}
\mathcal{T}_{33,\mathit{c}}^{\text{scalar}} 2\left( \frac{m}{2\pi L}\right) ^{\frac{D}{2%
}}\left[ \left( 1-D\right) \sum_{n=1}^{\infty } \frac{\cos(\theta \pi n)K_{\frac{D}{2}}\left( mnL\right)}{n^{\frac{D}{2}}}
 \right.  \\
\left. -mL\sum_{n=1}^{\infty }
\frac{\cos(\theta \pi n)K_{\frac{D}{2}-1}\left( mnL\right)}{n^{\frac{D}{2}-1}}
 \right] .
\label{Casimireffect-compactifiedspace}
\end{multline}%
The formula above corresponds to a general expression for the Casimir
pressure exerted by the vacuum fluctuations on the boundaries
 of the compactified manifold formed of two parallel planes separated by a length $L$. The
result presented in Eq.~(\ref{Casimireffect-compactifiedspace}) is the
quantum vacuum pressure for a massive scalar field submitted to
periodic boundary conditions in $D$ dimensions and agrees with
Refs.~\cite{Milton-2001, Ambjorn-Wolfram-Annals-of-Phys-1983}.

\begin{figure}
	\centering
	\includegraphics[width=0.7\linewidth]{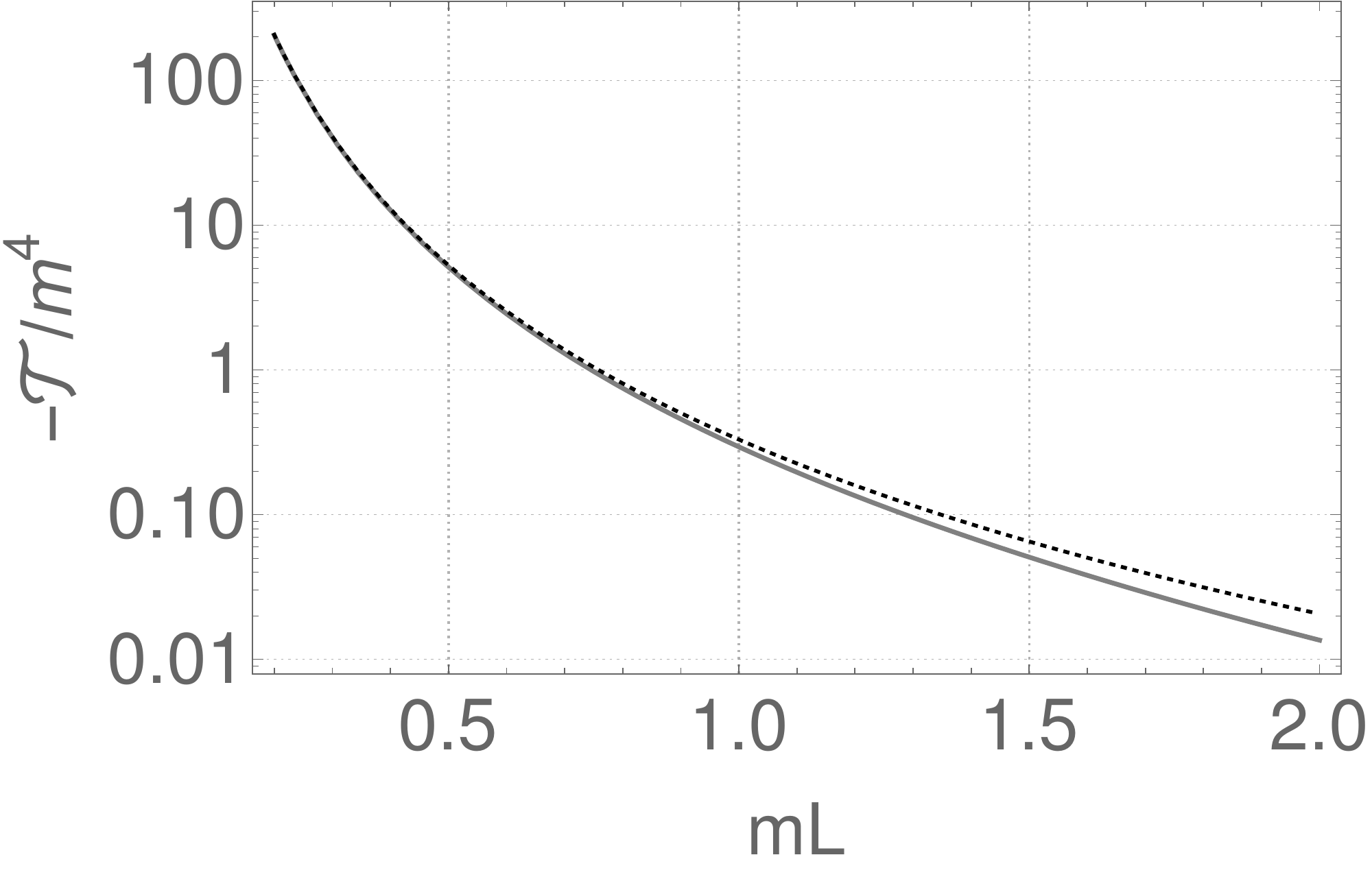}
	\caption{Casimir force as a function of the dimensionless product $mL$. Thick gray curve if the full expression, Eq~\eqref{Casimireffect-compactifiedspace-forDequals4} and the black dotted curve is the asymptotic formula for $mL\ll1$, Eq.~\eqref{eq:asymptotic-small-BesselK}. The Casimir force is always attractive in this scenario.}
	\label{fig:scalarlperiodic}
\end{figure}

For a $4$-dimensional Euclidean space, we obtain 
\begin{eqnarray}
\mathcal{T}_{33,\mathit{c}}^{\text{scalar}}\left( L,m\right) &=&-\frac{m^{2}}{2\pi
^{2}L^{2}}\left[ 3\sum_{n=1}^{\infty }\frac{\cos(\theta \pi n)}{n^{2}}\!K_{2}\left(
mnL\right)  \right.  \nonumber \\
&&\left. +mL\sum_{n=1}^{\infty }\frac{\cos(\theta \pi n)}{n}K_{1}\left( mnL\right) \right] .
\label{Casimireffect-compactifiedspace-forDequals4}
\end{eqnarray}%
From the following asymptotic formula of the Bessel function, 
\begin{equation}
K_{\nu }\left( z\right) \approx {2^{\nu -1}z^{-\nu }\Gamma \left( \nu
\right) },  \label{eq:asymptotic-small-BesselK}
\end{equation}%
evaluated for small values of its argument $\left( z\sim 0\right) $ and $%
\mathcal{R}$e$(\nu )>0$, we obtain the small-mass limit Casimir pressure ($%
mL\ll 1$) assuming periodic boundary conditions ($\theta=0$) 
\begin{equation}
\mathcal{T}_{33,\mathit{c}}^{\text{scalar}}\left( L,0\right) =-\frac{\pi ^{2}}{30L^{4}}\;,
\label{eq:CasimirEffectParticularCasePeriodicBC}
\end{equation}%
where we have neglected terms of $\mathcal{O}\left( m^{2}\right) $. The
vacuum fluctuation Casimir force per unit area is a finite negative
expression which suggests that the radiation pressure tends to diminish the
  distance $L$ between the planes. In Fig.~\ref{fig:scalarlperiodic} we plot both Eq.~\eqref{Casimireffect-compactifiedspace-forDequals4} and its asymptotic formula for $mL\ll1$, Eq.~\eqref{eq:asymptotic-small-BesselK}, which makes evident the agreement in the desired region.

An interesting comment is that the
corresponding negative Casimir pressure between two infinitely parallel
planes, when one imposes to the massless scalar field Dirichlet boundary
conditions, that is, $\phi (x_{3}=0)=\phi (x_{3}=L)=0$, is immediately
recovered when the plane separation distance $a$ is equal to the half
circumference length $L$ of the space dimension under compactification.

Furthermore, if we assume that we can build an antiperiodic boundary condition for one spatial direction, we get, instead of Eq.~\eqref{eq:CasimirEffectParticularCasePeriodicBC}, the scenario (with $\theta=1$),
\begin{equation}
\mathcal{T}_{33,\mathit{c}}^{\text{scalar}}\left( L,0\right) =\frac{7\pi ^{2}}{240L^{4}}\;,
\label{eq:CasimirEffectParticularCaseAntiPeriodicBC}
\end{equation}%
\noindent where the Casimir force is \textit{repulsive} instead of attractive. This flip from an attractive force to a repulsive force is due to the change in boundary conditions from periodic to antiperiodic (see Ref.~\cite{Flachi-etal-PRL-2017} and references therein). From a purely hypothetical point of view, one can notice that fine-tuning the parameter $\theta$ can smoothly change the strength of the Camisir force and then change its sign. One scenario where this could be justified is when considering a complex scalar field coupled to a constant gauge field along the restricted spatial direction, as discussed in Ref.~\cite{Cavalcanti-etal-IJMPA-2018} and references therein.

\section{Thermal Casimir pressure - scalar field}
\label{ThEf} 

In this section, thermal and boundary effects are treated simultaneously
through the generalized Matsubara prescription. We then consider a $D$%
-dimensional space with a double compactification ($d=2$) of the Euclidean
space corresponding to a compactified spatial dimension with length $L$ and
a compactification of the imaginary-time dimension with length $\beta $. In other words, we take the simultaneous compactification of both the $x_{0}$
and $x_{3}$ coordinates of the 4-vector $x$. 

Following the same steps as in the previous sections, the stress tensor
component $\mathcal{T}_{33,\mathit{c}}$ given by Eq.~(\ref%
{eq:previous_pressure_expression}) now becomes 
\begin{eqnarray}
\mathcal{T}_{33,\mathit{c}}^{\text{scalar}} &=&\frac{1}{2\beta L}\sum_{n_{1},n_{2}=-\infty
}^{+\infty }\int_{-\infty }^{\infty }\frac{d^{D-2}k_{\bot }}{\left( 2\pi
\right) ^{D-2}}  \nonumber \\
&&\times \left[ \frac{k_{n_{1}}^{2}-k_{n_{2}}^{2}-\left( k_{\bot
}^{2}+m^{2}\right) }{k_{n_{1}}^{2}+k_{n_{2}}^{2}+k_{\bot }^{2}+m^{2}}\right]
\!.\;\;\;
\end{eqnarray}%
Using dimensional regularization, Eqs.~(\ref%
{eq:firstmultidimensionalintegral}) and (\ref%
{eq:secondmultidimensionalintegral}), the previous formula is rewritten as 
\begin{multline}
\mathcal{T}_{33,\mathit{c}}^{\text{scalar}} =\Bigg\{ f_{s}\left( \nu ,\beta ,L\right) \times
\\
\left[ \sum_{n_{1},n_{2}=-\infty }^{+\infty }\frac{\left[
a_{1}\left(n_{1}+\theta_1/2\right)^{2}-a_{2}n_{2}^{2}-c^{2}\right] \Gamma \left( \nu \right) }{\left[
a_{1}\left(n_{1}+\theta_1/2\right)^{2}+a_{2}n_{2}^{2}+c^{2}\right] ^{\nu }}\right.   
\\
\;\; \left. -\sum_{n_{1},n_{2}=-\infty }^{+\infty }\frac{\left(
s-\nu \right) \Gamma \left( \nu -1\right) }{\left[
a_{1}\left(n_{1}+\theta_1/2\right)^{2}+a_{2}n_{2}^{2}+c^{2}\right] ^{\nu -1}}\right] \Bigg\}
_{\!\!s=1}\;\;,  \label{eq:tensor_pre_summation2}
\end{multline}%
where $a_{1}=L^{-2}$, $a_{2}=\beta ^{-2}$, $c=m/{2\pi }$, $\nu =s-\left(
D-2\right) /2$, and $f_{s}\left( \nu ,\beta ,L\right) $ is a function given
by 
\begin{equation}
f_{s}\left( \nu ,\beta ,L\right) =\frac{1}{2\beta L}\frac{1}{\left( 4\pi
\right) ^{s-\nu }\left( 2\pi \right) ^{2\left( \nu -1\right) }\Gamma \left(
s\right) }.  \label{DefinitionOfTheGeneralFunctionF}
\end{equation}%
As before, $\theta_1$ is a parameter that controls whether we have periodic ($\theta_1=0$) or antiperiodic ($\theta_1=1$) boundary conditions in the spatial direction. Notice that $\theta_2$ is absent in the expression because the boundary condition in the imaginary time is restricted, due to the KMS condition~\cite{Khanna-et-al-Thermal-Book-2009}, to be periodic ($\theta_2=0$) for bosons. 
Adding and subtracting the term $\left( a_{2}n_{2}^{2}+c^{2}\right) \Gamma
\left( \nu \right) $ in the numerator of the first term on the right-hand
side of Eq.~(\ref{eq:tensor_pre_summation2}), we obtain 
\begin{eqnarray}
\mathcal{T}_{33,\mathit{c}}^{\text{scalar}} &=&\left\{ f_{s}\left( \nu ,\beta ,L\right)
\Gamma \left( \nu -1\right) \Big[ \left( 2\nu -s-1\right) \right.  
\nonumber \\
&&\times Z_{2}^{c^{2}}\!\!\left( \nu -1;a_{1},a_{2};\theta_1\right) -2c^{2}\left( \nu
-1\right) Z_{2}^{c^{2}}\!\!\left( \nu ;a_{1},a_{2};\theta_1\right)   \nonumber \\
&&\left. \left. +2a_{2}\frac{\partial }{\partial a_{2}}Z_{2}^{c^{2}}\!\!%
\left( \nu -1;a_{1},a_{2};\theta_1\right) \right] \right\} _{s=1},
\label{eq:tensor_in_terms_of_Zeta}
\end{eqnarray}%
where we have used the definition of the two-dimensional Epstein--Hurwitz
zeta function, $Z_{2}^{c^{2}}\!\!\left( \nu ;a_{1},a_{2};\theta_1\right) $, obtained
from Eq.~(\ref{eq:multidimensional_Epstein_Hurwitz_zeta_function}) for $d=2$%
. From Eq.~(\ref{eq:analytical_continuation_of_MultiZeta}), we get for $d=2$ 
\begin{eqnarray}
Z_{2}^{c^{2}}\!\left( \nu ;a_{1},a_{2}\right)  &=&\frac{2\pi }{\sqrt{%
a_{1}a_{2}}\;\Gamma \left( \nu \right) }\left[ \frac{1}{2c^{2\left( \nu
-1\right) }}\Gamma \left( \nu -1\right) \right.   \nonumber \\
&&+2\sum_{n_{1}=1}^{\infty }\left( \frac{\pi n_{1}}{c\sqrt{a_{1}}}\right)
^{\nu -1}\cos(\theta_1\pi n_1)\nonumber\\&&\times K_{\nu -1}\left( 2\pi c\frac{n_{1}}{\sqrt{a_{1}}}\right)   \nonumber \\
&&+2\sum_{n_{2}=1}^{\infty }\left( \frac{\pi n_{2}}{c\sqrt{a_{2}}}\right)
^{\nu -1}K_{\nu -1}\left( 2\pi c\frac{n_{2}}{\sqrt{a_{2}}}\right)   \nonumber \\
&&+2^{2}\sum_{n_{1},n_{2}=1}^{\infty }\left( \frac{\pi }{c}\sqrt{\frac{%
n_{1}^{2}}{a_{1}}+\frac{n_{2}^{2}}{a_{2}}}\right) ^{\nu -1}  \cos(\theta_1\pi n_1)\nonumber \\
&&\left. \times K_{\nu -1}\left( 2{\pi }{c}\sqrt{\frac{n_{1}^{2}}{a_{1}}+%
\frac{n_{2}^{2}}{a_{2}}}\right) \right] .
\label{eq:analytical_continuation_of_BiZeta}
\end{eqnarray}%
Substituting Eq.~(\ref{eq:analytical_continuation_of_BiZeta}) in Eq.~(\ref%
{eq:tensor_in_terms_of_Zeta}), splitting $\mathcal{T}_{33,\mathit{c}}^{\text{scalar}}$
into three terms, $\mathcal{T}_{33,\mathit{c}}^{\text{scalar}}=\mathcal{T}_{n_1}^{\text{scalar}}+\mathcal{T}_{n_{2}}^{\text{scalar}}+\mathcal{T}_{n_{1}n_{2}}^{\text{scalar}},$ after removing nonphysical terms, we
have 
\begin{eqnarray}
\mathcal{T}_{n_{1}}^{\text{scalar}} &=&\frac{4\pi }{\sqrt{a_{1}a_{2}}}%
f_{s}\left( \nu ,\beta ,L\right) \Bigg[\left( 2\nu -s-2\right)
\sum_{n_{1}=1}^{\infty }\left( \frac{\pi n_{1}}{c\sqrt{a_{1}}}\right) ^{\nu
-2}  \nonumber \\
&&\times \cos(\theta_1\pi n_1)K_{\nu -2}\left( 2\pi c\frac{n_{1}}{\sqrt{a_{1}}}\right)
 \nonumber \\
&& -2c^{2}\sum_{n_{1}=1}^{\infty }\left( \frac{\pi n_{1}}{c\sqrt{a_{1}}}\right)
^{\nu -1} \cos(\theta_1\pi n_1) \nonumber \\
&& \times K_{\nu -1}\left( 2\pi c\frac{n_{1}}{\sqrt{a_{1}}}%
\right) \Bigg] \Bigg\vert _{s=1},
\end{eqnarray}%
which corresponds to the contribution to the Casimir pressure due to vacuum
fluctuations only. Using the definition (\ref%
{DefinitionOfTheGeneralFunctionF}), for $a_{1}=L^{-2}$, $a_{2}=\beta ^{-2}$, 
$c=m/{2\pi }$, $\nu =s-\left( D-2\right) /2$, the result obtained for Eq.~(\ref%
{Casimireffect-compactifiedspace}), shown in the previous section, is recovered.

Also, 
\begin{eqnarray}
\mathcal{T}_{n_{2}}^{\text{scalar}} &=&\frac{4\pi }{\sqrt{a_{1}a_{2}}}%
f_{s}(\nu ,\beta ,L)\left[ \left( 2\nu -s-2\right) \right.  \nonumber \\
&&\times \sum_{n_{2}=1}^{\infty }\left( \frac{\pi n_{2}}{c\sqrt{a_{2}}}%
\right) ^{\nu -2}K_{\nu -2}\left( 2\pi c\frac{n_{2}}{\sqrt{a_{2}}}\right) 
\nonumber \\
&&-2c^{2}\sum_{n_{2}=1}^{\infty }\left( \frac{\pi n_{2}}{c\sqrt{a_{2}}}%
\right) ^{\nu -1}K_{\nu -1}\left( 2\pi c\frac{n_{2}}{\sqrt{a_{2}}}\right) 
\nonumber \\
&&\left. \left. +2a_{2}\frac{\partial }{\partial a_{2}}\sum_{n_{2}=1}^{%
\infty }\left( \frac{\pi n_{2}}{c\sqrt{a_{2}}}\right) ^{\nu -2}K_{\nu
-2}\left( 2\pi c\frac{n_{2}}{\sqrt{a_{2}}}\right) \right] \right\vert _{s=1},
\nonumber \\
&&
\end{eqnarray}%
yields 
\begin{equation}
\mathcal{T}_{n_{2}}^{\text{scalar}}\left( \beta ,m\right) =2\left( 
\frac{m}{2\pi \beta }\right) ^{\frac{D}{2}}\!\sum_{n_{2}=1}^{\infty
}\!\left( \frac{1}{n_{2}}\right) ^{\frac{D}{2}}\!\!K_{\frac{D}{2}}\left(
m\beta n_{2}\right) ,  \label{eq:PureThermalCasimireffectformula}
\end{equation}%
which is the Casimir force formula due exclusively to the thermal
fluctuations. The final form of Eq.~(\ref{eq:PureThermalCasimireffectformula}%
) was obtained by means of the useful recurrence formula for Bessel
functions, 
\begin{equation}
K_{\alpha -1}\left( z\right) -K_{\alpha +1}\left( z\right) =-\frac{2\alpha }{%
z}K_{\alpha }\left( z\right) .  \label{eq:recurrenceBesselKformula}
\end{equation}%
For $D=4$, we find 
\begin{equation}
\mathcal{T}_{n_{2}}^{\text{scalar}}\left( \beta ,m\right) =\left( 
\frac{m^{2}}{2\pi ^{2}\beta ^{2}}\right) \sum_{n_{2}=1}^{\infty }\left( 
\frac{1}{n_{2}}\right) ^{2}\!K_{2}\left( m\beta n_{2}\right) .
\label{eq:PureThermalCasimireffectformulaOnceAgain}
\end{equation}%
Using Eq.~(\ref{eq:asymptotic-small-BesselK}), we obtain the small-mass
limit for the purely thermal Casimir pressure ($m\beta \ll 1$) 
\begin{equation}
\mathcal{T}_{n_{2}}^{\text{scalar}}\left( \beta ,0\right) =\frac{\pi
^{2}}{90\beta ^{4}},
\label{eq:PureThermalCasimireffectformulaOnceAgainandAgain}
\end{equation}%
which is in accordance with the well-known Stefan-Bolztmann thermal
radiation pressure result. This is a finite positive force per unit area
which is more intense than vacuum radiation Casimir pressure for low values
of $\beta $ (high-temperature or classical limit).

Finally, the formula 
\begin{eqnarray}
\mathcal{T}_{n_{1}n_{2}}^{\text{scalar}} &=&\frac{8\pi }{\sqrt{a_{1}a_{2}}}%
f_{s}\left( \nu ,\beta ,L\right)  \nonumber \\
&&\times \left\{ \left( 2\nu -s-2\right) \sum_{n_{1},n_{2}=1}^{\infty
}\left( \frac{\pi }{c}\sqrt{\frac{n_{1}^{2}}{a_{1}}+\frac{n_{2}^{2}}{a_{2}}}%
\right) ^{\nu -2}\right.  \nonumber \\
&&\times \cos(\theta_1 \pi n_1)K_{\nu -2}\left( 2\pi c\sqrt{\frac{n_{1}^{2}}{a_{1}}+\frac{n_{2}^{2}%
}{a_{2}}}\right)  \nonumber \\
&&-2c^{2}\sum_{n_{1},n_{2}=1}^{\infty }\left( \frac{\pi }{c}\sqrt{\frac{%
n_{1}^{2}}{a_{1}}+\frac{n_{2}^{2}}{a_{2}}}\right) ^{\nu -1}  \nonumber \\
&&\times \cos(\theta_1\pi n_1)K_{\nu -1}\left( 2\pi c\sqrt{\frac{n_{1}^{2}}{a_{1}}+\frac{n_{2}^{2}%
}{a_{2}}}\right)  \nonumber \\
&&+2a_{2}\frac{\partial }{\partial a_{2}}\sum_{n_{1},n_{2}=1}^{\infty }\left[
\left( \frac{\pi }{c}\sqrt{\frac{n_{1}^{2}}{a_{1}}+\frac{n_{2}^{2}}{a_{2}}}%
\right) ^{\nu -2} \right.  \nonumber \\
&&\left. \left. \left. \times \cos(\theta_1\pi n_1)K_{\nu -2}\left( 2\pi c\sqrt{\frac{n_{1}^{2}}{%
a_{1}}+\frac{n_{2}^{2}}{a_{2}}}\right) \right] \right\} \right\vert _{s=1},
\end{eqnarray}%
or 
\begin{eqnarray}
\mathcal{T}_{n_{1}n_{2}}^{\text{scalar}}\!\left( L,\beta ,m\right) \!
&=&\!4\left( \frac{m}{2\pi }\right) ^{\frac{D}{2}}\!\!\left[
\sum_{n_{1},n_{2}=1}^{\infty }\!\left( \frac{1}{\sqrt{%
n_{1}^{2}L^{2}+n_{2}^{2}\beta ^{2}}}\right) ^{\frac{D}{2}}\right.  \nonumber \\
&&\times \left( \frac{\left( 1-D\right) n_{1}^{2}L^{2}+n_{2}^{2}\beta ^{2}}{%
n_{1}^{2}L^{2}+n_{2}^{2}\beta ^{2}}\right) \cos(\theta_1\pi n_1)  \nonumber \\
&&\times K_{\frac{D}{2}}\left( m\sqrt{n_{1}^{2}L^{2}+n_{2}^{2}\beta ^{2}}%
\right)  \nonumber \\
&&{-}m\sum_{n_{1},n_{2}=1}^{\infty }\!n_{1}^{2}L^{2}\!\left( \frac{1}{\sqrt{%
n_{1}^{2}L^{2}+n_{2}^{2}\beta ^{2}}}\right) ^{\frac{D}{2}+1}  \nonumber \\
&&\times \cos(\theta_1\pi n_1) \nonumber \\
&&\left. \times K_{\frac{D}{2}-1}\left( m\sqrt{n_{1}^{2}L^{2}+n_{2}^{2}\beta
^{2}}\right) \right] ,
\end{eqnarray}%
gives the corrections to the Casimir pressure in a compact space in the
presence of a massive scalar field in a thermal bath at temperature $1/\beta $. In
order to obtain the final form of the above expression, we have used the
recurrence formula given by Eq.~(\ref{eq:recurrenceBesselKformula}).
Considering $D=4$, we get 
\begin{eqnarray}
\mathcal{T}_{n_{1}n_{2}}^{\text{scalar}}\!\left( L,\beta ,m\right) \!
&=&-\left( \frac{m}{\pi }\right) ^{2}\left[ \sum_{n_{1},n_{2}=1}^{\infty }%
\frac{3n_{1}^{2}L^{2}-n_{2}^{2}\beta ^{2}}{\left(
n_{1}^{2}L^{2}+n_{2}^{2}\beta ^{2}\right) ^{2}}\right.  \nonumber \\
&&\times \cos(\theta_1\pi n_1) K_{2}\left( m\sqrt{n_{1}^{2}L^{2}+n_{2}^{2}\beta ^{2}}\right) 
\nonumber \\
&&{+}m\sum_{n_{1},n_{2}=1}^{\infty }\frac{n_{1}^{2}L^{2}}{\left(
n_{1}^{2}L^{2}+n_{2}^{2}\beta ^{2}\right) ^{\frac{3}{2}}}  \cos(\theta_1\pi n_1)\nonumber \\
&&\left. \times K_{1}\left( m\sqrt{n_{1}^{2}L^{2}+n_{2}^{2}\beta ^{2}}%
\right) \right] ,
\end{eqnarray}%
which is valid for arbitrary values of $m$, $L$ and $\beta $. Using Eq.~(\ref%
{eq:asymptotic-small-BesselK}), we can show that in the small-mass case it
reduces to 
\begin{equation}
\mathcal{T}_{n_{1}n_{2}}^{\text{scalar}}\!\left( L,\beta ,0\right) \!=-\frac{2}{%
\pi ^{2}}\sum_{n_{1},n_{2}=1}^{\infty }\cos(\theta_1\pi n_1)\frac{3n_{1}^{2}L^{2}-n_{2}^{2}\beta
^{2}}{\left( n_{1}^{2}L^{2}+n_{2}^{2}\beta ^{2}\right) ^{3}},
\label{eq:vacuumthermalpressuresmallmasslesslimit4dimensions}
\end{equation}%
where we have disregarded terms of $\mathcal{O}\left( m^{2}\right) $. The
corresponding expression for Dirichlet boundary conditions can be obtained
by substituting $L=2a$.

Now we can go back and show asymptotic formula for $\mathcal{T}_{33,c}$ in the small-mass scenario making use of the expressions for $\mathcal{T}_{n_{1}}$ (Eq.~\eqref{Casimireffect-compactifiedspace-forDequals4}), $\mathcal{T}_{n_{2}}$ (Eq.~\eqref{eq:PureThermalCasimireffectformulaOnceAgainandAgain}) and $\mathcal{T}_{n_{1}n_{2}}$ (Eq.~\eqref{eq:vacuumthermalpressuresmallmasslesslimit4dimensions}). For periodic boundary conditions in space the small-mass limit of the Casimir force is

\begin{equation}
\mathcal{T}_{33,c}^{\text{scalar}}(L,\beta,0) = 
-\frac{\pi ^{2}}{30L^{4}} +\frac{\pi^{2}}{90\beta ^{4}} 
-\frac{2}{\pi ^{2}}\sum_{n_{1},n_{2}=1}^{\infty }\frac{3n_{1}^{2}L^{2}-n_{2}^{2}\beta^{2}}{\left( n_{1}^{2}L^{2}+n_{2}^{2}\beta ^{2}\right) ^{3}}
\label{eq:CasimirForce-BetaL-Periodic}.
\end{equation}

\begin{figure}
	\centering
	\includegraphics[width=0.7\linewidth]{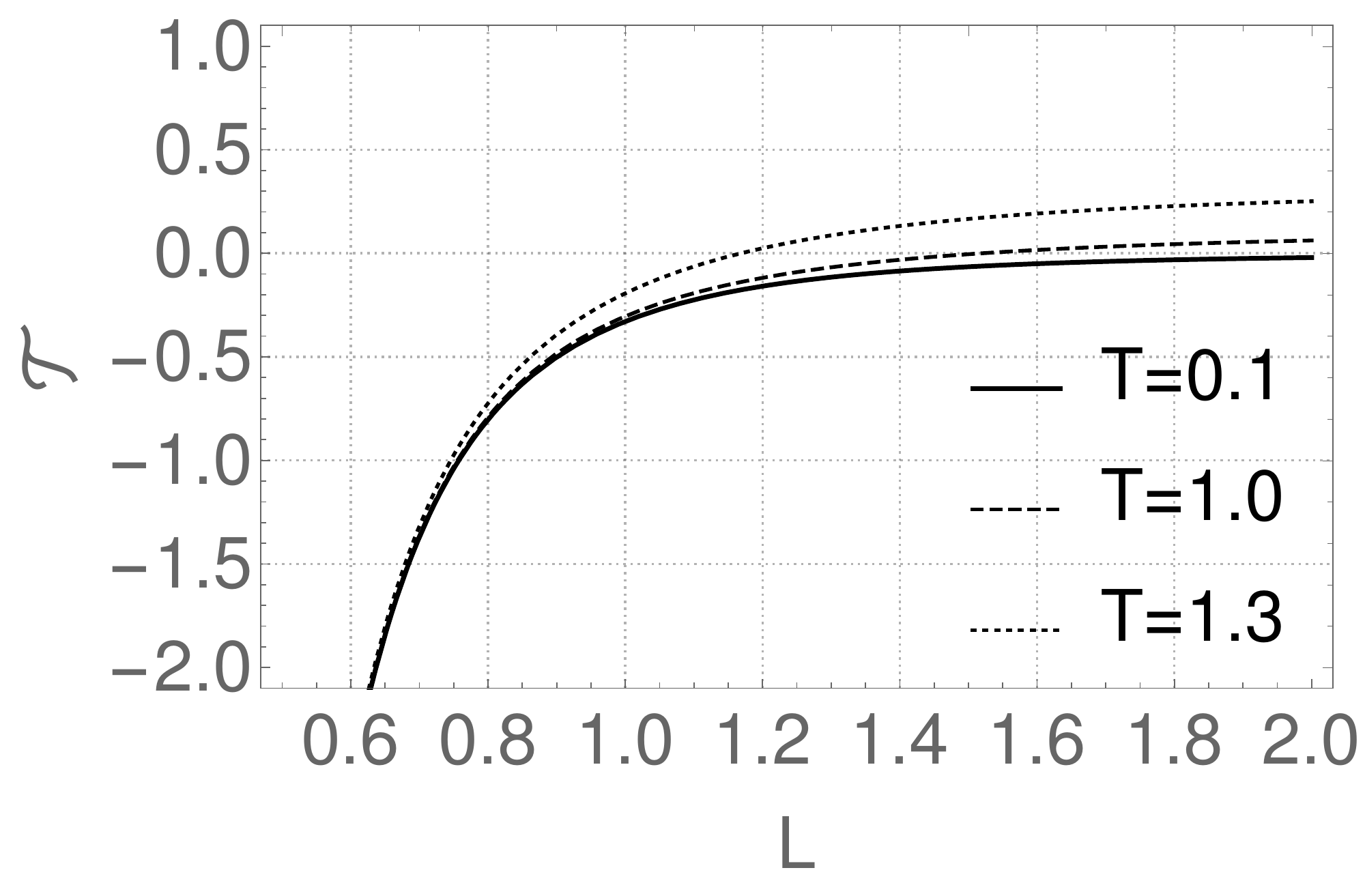}
	\caption{Strength of the force with respect to the size of the cavity for different heat bath temperatures. With the increase of the temperature, we get a characteristic length above which the force becomes repulsive.}
	\label{fig:scalarcurveltperiodic}
\end{figure}

\begin{figure}
	\centering
	\includegraphics[width=0.7\linewidth]{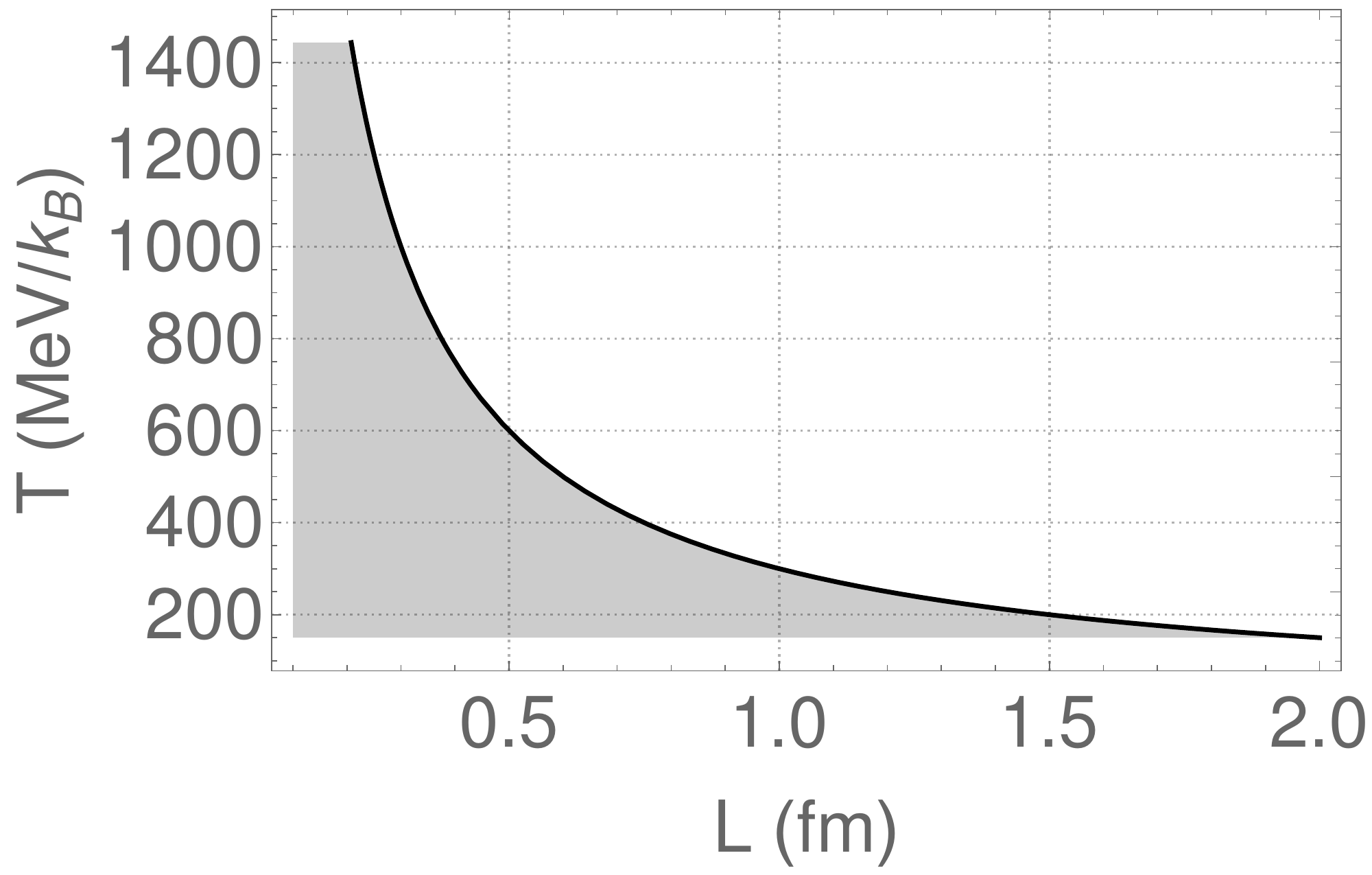}
	\caption{Taking the scalar field with periodic boundary conditions, the diagram shows the transition from an attractive (filled region) to a repulsive (unfilled region) Casimir force.}
	\label{fig:scalarphasediagram}
\end{figure}

Notice that while the vaccum fluctuation between the plates produces an attractive behavior, the thermal fluctuation produces a repulsive behavior that tries to compensate it. Let us rewrite it introducing dimensionaless function $g(\xi)$, where $\xi = L/\beta = LT$, 
\begin{equation}
\mathcal{T}_{33,c}^{\text{scalar}}(L,\beta,0) = 
-\frac{\pi ^{2}}{30L^{4}} g^{\text{scalar}}_{\theta=0}(L/\beta)
\label{eq:CasimirForce-BetaL-Periodic-Adimensional}.
\end{equation}
\begin{figure}
	\centering
	\includegraphics[width=0.7\linewidth]{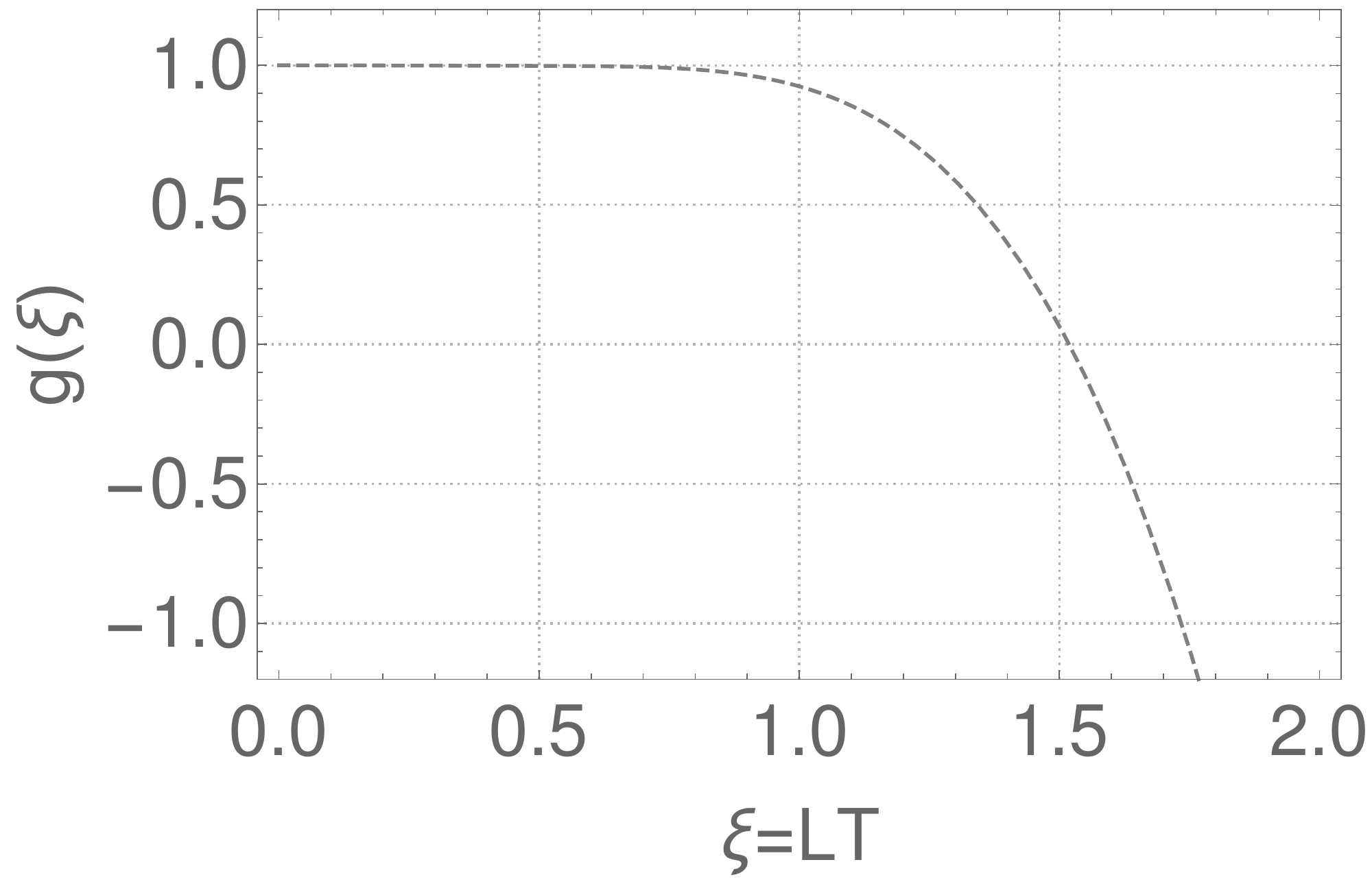}
	\caption{Behavior of the function $g(\xi)$, for scalar fields, using periodic boundary conditions. Curve plotted using 500 terms of the double sum. The root of the function, $g(\xi^*)=0$, is around $\xi \sim 1.52 \hbar c/k_B = 300/k_B \text{fm} \cdot \text{MeV}$ (back into SI units).}
	\label{fig:scalarltperiodic}
\end{figure}
\noindent In our scenario the expression for $g(\xi)$ is
\begin{equation}
g^{\text{scalar}}_{\theta=0}(\xi) = 1- \frac{\xi^4}{3} 
+\frac{60}{\pi ^{4}}\sum_{n_{1},n_{2}=1}^{\infty }\frac{3n_{1}^{2}-n_{2}^{2}/\xi^{2}}{\left( n_{1}^{2}+n_{2}^{2}/\xi ^{2}\right) ^{3}},
\end{equation}
\noindent and its behavior is exhibited in Fig.~\ref{fig:scalarltperiodic} for small values of $\xi$. It is evident that as the parameter $\xi$ increase the value of the function $g$ gets smaller and at some particular value $\xi \simeq 1.52$ it changes sign. That is, the effective Casimir forces -- considering the vacuum and thermal contribution -- became repulsive instead of attractive. 
We show the behavior of the force in Fig.~\ref{fig:scalarcurveltperiodic} for different choices of the heat bath temperature ($T=0.1, 1.0, 1.3$). This illustrates that a change in behavior from an attractive to a repulsive force occurs at some characteristic length and that the higher the temperature, the smaller the cavity size at which this occurs.
The flip in the Casimir fore occurs for the particular value of $T \simeq 1.52/L$, meaning that the required heat bath temperature to turn the effective force repulsive depends on the characteristic length $L$ of the system under study. 
Back into SI units, $LT \simeq 1.52 \hbar c/k_B = 3.49 \times 10^{12} \text{fm} \cdot \text{K} = 300/k_B\, \text{fm} \cdot \text{MeV} $. Meaning that to a cavity around the typical size of a meson (1 fm) the temperature at which the Casimir force vanishes is around $3.49\times10^{12}$K (or $300$MeV). To illustrate, we exhibit the `phase diagram' for the scalar field under periodic boundary conditions, the filled region indicates an attractive force and the unfilled region indicates a repulsive force, see Fig.~\ref{fig:scalarphasediagram}.

In the scenario with antiperiodic boundary condition the small-mass limit of the Casimir force is, instead of Eq.~\eqref{eq:CasimirForce-BetaL-Periodic}, given by
\begin{equation}
\mathcal{T}_{33,c}^{\text{scalar}}(L,\beta,0) = 
\frac{7\pi ^{2}}{240L^{4}}g^{\text{scalar}}_{\theta=1}(L/\beta),
\label{eq:CasimirForce-BetaL-AntiPeriodic}
\end{equation}
\noindent where the expression for $g^{\text{scalar}}_{\theta=1}(\xi)$ is
\begin{equation}
g^{\text{scalar}}_{\theta=1}(\xi) = 1 +\frac{8\xi^4}{21} 
-\frac{480}{7\pi ^{4}}\sum_{n_{1},n_{2}=1}^{\infty }(-1)^{n_1}\frac{3n_{1}^{2}-n_{2}^{2}/\xi^{2}}{\left( n_{1}^{2}+n_{2}^{2}/\xi ^{2}\right) ^{3}}.
\end{equation}
\noindent For antiperiodic boundary condition the Casimir force is already repulsive, and the contribution from the dimensionaless function $g(\xi)$ is positive (the double sum is not enought to flip the sign). Meaning that it stays a repulsive force regardless of the thermal contribution. This is expected, the contribution for the increase in the heat bath is not to ``flip" the sign of the Casimir force, but instead to introduce a repulsive behavior. If the boundary conditions already imposes a repulsive force, the increase in temperature is only able to intensify it.

%
\section{Thermal Casimir pressure - fermionic field}
\label{ThEf_Fermion} 

The procedure to compute the quantum vacuum pressure for fermions is similar to what we just exhibited for scalar fields as can be noticed comparing Eq.~\eqref{eq:previous_pressure_expression_fermion} and Eq.~\eqref{eq:previous_pressure_expression}. If we follow the path from Sec.~\ref{VacuumPressureInToroidalSpace} the bulk expression Eq.~\eqref{eq:previous_pressure_expression_fermion} is changed into
\begin{equation}
\mathcal{T}_{33,\mathit{c}}^{\text{fermion}}=\!\frac{1}{2L}\!\sum_{n=-\infty }^{+\infty
}\!\int_{-\infty }^{\infty }\frac{d^{D-1}k_{\bot }}{\left( 2\pi \right)
	^{D-1}}\!\left[ \frac{k_{n}^{2}-\left( k_{\bot }^{2}+m^{2}\right) }{%
	k_{n}^{2}+k_{\bot }^{2}+m^{2}}\right] \!.\;\;\;
\label{eq:tensor_matsubarizado_espacialmente_fermion}
\end{equation}
\noindent Then, we integrate over the orthogonal momenta $k_\bot$ and use the general definition of the multidimensional Epstein-Hurwitz zeta function 
\begin{eqnarray}
\mathcal{T}_{33,\mathit{c}}^{\text{fermion}} &=&2^{1+\lfloor D/2\rfloor}\Big\{ f_{s}\left( \nu ,L\right) \Gamma\left(\nu\right) 
\nonumber\\&&
\times\left[
 Z_{1}^{c^{2}}\!\left( \nu -1;a;\theta\right)  - c^{2} Z_{1}^{c^{2}}\!\left( \nu
;a;\theta\right) \right] \Big\} _{\!\!s=1}.
\label{eq:tensor_more_than_pre_summation2_fermion}
\end{eqnarray}%
\noindent Afterwards, we extend the zeta function to the whole complex plane and obtain the expression for the pressure in a cavity (one restricted spatial dimension) for a $D$ dimensional Euclidian space-time,
\begin{multline}
\mathcal{T}_{33,\mathit{c}}^{\text{fermion}} =2^{1+\lfloor D/2\rfloor}\left( \frac{m}{2\pi L}\right) ^{\frac{D}{2%
}}\Bigg[\\ \left( 1-D\right) \sum_{n=1}^{\infty } \frac{\cos(\theta \pi n)K_{\frac{D}{2}}\left( mnL\right)}{n^{\frac{D}{2}}}
\\-mL\sum_{n=1}^{\infty }
\frac{\cos(\theta \pi n)K_{\frac{D}{2}-1}\left( mnL\right)}{n^{\frac{D}{2}-1}}
\Bigg] .
\label{Casimireffect-compactifiedspace_fermion}
\end{multline}

The asymptotic expression in the small-mass limit ($mL\ll1$) is
\begin{subequations}
	\begin{eqnarray}
\mathcal{T}_{33,\mathit{c}}^{\text{fermion}}\left( L,0\right) \sim-\frac{2\pi ^{2}}{15L^{4}}\;,
&&\quad \theta = 0\label{eq:CasimirEffectParticularCasePeriodicBC_fermion};
\\
\mathcal{T}_{33,\mathit{c}}^{\text{fermion}}\left( L,0\right) \sim\frac{7\pi ^{2}}{60L^{4}}\;,
&&\quad\theta = 1.\label{eq:CasimirEffectParticularCaseAntiPeriodicBC_fermion}
	\end{eqnarray}
\end{subequations}

Showing that the pressure for the fermionic field behaves just like the scalar scenario if we consider the same boundary conditions. Indeed, if we compare the pressure in a cavity of size $L$ assuming the general case of $D$ dimensions, the ratio between the pressure for a fermionic field, Eq.~\eqref{Casimireffect-compactifiedspace_fermion}, and the pressure for a scalar field, Eq.~\eqref{Casimireffect-compactifiedspace}, is just number independent of any characteristics of the model,
\begin{equation}
\frac{\mathcal{T}_{33,\mathit{c}}^{\text{fermion}}}{\mathcal{T}_{33,\mathit{c}}^{\text{scalar}}} = 2^{\lfloor D/2\rfloor}.
\end{equation}
\noindent This result suggests, once again, that the choice of boundary conditions is more significant than the nature of the quantum field when we deal with Casimir forces.

The scenario discussed in Sec.~\ref{ThEf}, where thermal and boundary effects are considered, is a bit different from the scalar field. This occurs because the choice of boundary condition in the imaginary-time direction is restricted by the KMS condition. Therefore, we might expect from the beginning some difference between fermions, which use antiperiodic boundary conditions, and bosons, which use periodic boundary conditions.

In the fermionic scenario, the expression from Eq.~(\ref%
{eq:tensor_pre_summation2}) becames
\begin{multline}
\mathcal{T}_{33,\mathit{c}}^{\text{fermion}} =\Bigg\{ 2^{\lfloor D/2 \rfloor+1}f_{s}\left( \nu ,\beta ,L\right) \Gamma \left( \nu \right) \times
\\
\sum_{n_{1},n_{2}=-\infty }^{+\infty }\frac{
	a_{1}\left(n_{1}+\theta_1/2\right)^{2}}{\left[
	a_{1}\left(n_{1}+\theta_1/2\right)^{2}+a_{2}\left(n_{2}+\frac{1}{2}\right)^{2}+c^{2}\right] ^{\nu }} \Bigg\}
_{\!\!s=1}\;\;,  \label{eq:tensor_pre_summation2_fermion}
\end{multline}%

After 

\begin{eqnarray}
\mathcal{T}_{33,\mathit{c}}^{\text{scalar}} &=&\left\{ 2^{\lfloor D/2 \rfloor+1}f_{s}\left( \nu ,\beta ,L\right)
\Gamma \left( \nu\right) \Big[\right.  
\nonumber \\
&& Z_{2}^{c^{2}}\!\!\left( \nu -1;a_{1},a_{2};\theta_1,\frac{1}{2}\right) -c^{2} Z_{2}^{c^{2}}\!\!\left( \nu ;a_{1},a_{2};\theta_1,\frac{1}{2}\right)   \nonumber \\
&&\left. \left. +\frac{a_{2}}{\nu-1}\frac{\partial }{\partial a_{2}}Z_{2}^{c^{2}}\!\!%
\left( \nu -1;a_{1},a_{2};\theta_1,\frac{1}{2}\right) \right] \right\} _{s=1},
\label{eq:tensor_in_terms_of_Zeta_fermion}
\end{eqnarray}%

If we compare this expression with Eq.~\eqref{eq:tensor_in_terms_of_Zeta_fermion} we see that the only difference (for $s=1$) is a global multiplicative factor $2^{\lfloor D/2 \rfloor}$ and the fact that $\theta_2=1/2$, meaning the presence of the antiperiodic boundary condition for the imaginary-time direction. Following the same decomposition as the previous section, that is, $\mathcal{T}_{33,\mathit{c}}=\mathcal{T}_{n_1}+\mathcal{T}_{n_{2}}+\mathcal{T}_{n_{1}n_{2}},$ we get for each component
\begin{subequations}
\begin{multline}
\mathcal{T}_{n_1}^{\text{fermionic}} = 
2^{1+\lfloor D/2\rfloor}\left( \frac{m}{2\pi L}\right) ^{\frac{D}{2%
}}\\\times\Bigg[ \left( 1-D\right) \sum_{n_1=1}^{\infty } \frac{\cos(\theta_1 \pi n_1)K_{\frac{D}{2}}\left( mn_1 L\right)}{n_1^{\frac{D}{2}}}
\\-mL\sum_{n_1=1}^{\infty }
\frac{\cos(\theta_1 \pi n_1)K_{\frac{D}{2}-1}\left( mn_1 L\right)}{n_1^{\frac{D}{2}-1}}
\Bigg];
\end{multline}
\begin{multline}
\mathcal{T}_{n_2}^{\text{fermionic}} = 
2^{1+\lfloor D/2\rfloor}
\left( 
\frac{m}{2\pi \beta }\right) ^{\frac{D}{2}}\!\sum_{n_{2}=1}^{\infty
}\!(-1)^{n_2} \frac{K_{\frac{D}{2}}\left(
m\beta n_{2}\right)}{n_{2}^{\frac{D}{2}}} ;
\end{multline}
\begin{multline}
\mathcal{T}_{n_3}^{\text{fermionic}} = 2^{2+\lfloor D/2\rfloor}\left( \frac{m}{2\pi }\right) ^{\frac{D}{2}}\!\!\left[
\sum_{n_{1},n_{2}=1}^{\infty }(-1)^{n_2}\!\right.  
\\ \times \left( \frac{\left( 1-D\right) n_{1}^{2}L^{2}+n_{2}^{2}\beta ^{2}}{%
	\left(n_{1}^{2}L^{2}+n_{2}^{2}\beta ^{2}\right)^{\frac{D}{4}+1}}\right) \cos(\theta_1\pi n_1)  
\\\times K_{\frac{D}{2}}\left( m\sqrt{n_{1}^{2}L^{2}+n_{2}^{2}\beta ^{2}}%
\right)  
\\{-}m\sum_{n_{1},n_{2}=1}^{\infty }\!\!(-1)^{n_2}\left( \frac{n_{1}^{2}L^{2}}{\left(%
		n_{1}^{2}L^{2}+n_{2}^{2}\beta ^{2}\right)^{\frac{D}{4}+\frac{1}{2}}}\right)   
\\ \times \cos(\theta_1\pi n_1) 
\left.  K_{\frac{D}{2}-1}\left( m\sqrt{n_{1}^{2}L^{2}+n_{2}^{2}\beta
	^{2}}\right) \right] .
\end{multline}
\end{subequations}

In the small-mass limit and taking periodic boundary conditions in space this simplifies to
\begin{subequations}
\begin{equation}
\mathcal{T}_{33,c}^{\text{fermionic}}(L,\beta,0) = 
-\frac{2\pi ^{2}}{15L^{4}} g^{\text{fermionic}}_{\theta=0} (LT)
\label{eq:CasimirForce-BetaL-Periodic_fermion}.
\end{equation}
\begin{multline}
g^{\text{fermionic}}_{\theta=0} (\xi) = 1
+\frac{7\xi^4}{24} \\
+\frac{60}{\pi ^{4}}\sum_{n_{1},n_{2}=1}^{\infty }(-1)^{n_2}\frac{3n_{1}^{2}-n_{2}^{2}/\xi^{2}}{\left( n_{1}^{2}+n_{2}^{2}/\xi^{2}\right) ^{3}}
\end{multline}
While for antiperiodic boundary conditions in space this becames
\begin{equation}
\mathcal{T}_{33,c}^{\text{fermionic}}(L,\beta,0) = 
\frac{7\pi ^{2}}{60L^{4}} g^{\text{fermionic}}_{\theta=1} (LT)
\label{eq:CasimirForce-BetaL-AntiPeriodic_fermion}.
\end{equation}
\begin{multline}
g^{\text{fermionic}}_{\theta=1} (\xi) = 
1 -\frac{\xi^4}{3} 
\\-\frac{480}{7\pi ^{4}}\sum_{n_{1},n_{2}=1}^{\infty }(-1)^{n_1+n_2}\frac{3n_{1}^{2}-n_{2}^{2}/\xi^{2}}{\left( n_{1}^{2}+n_{2}^{2}/\xi ^{2}\right) ^{3}}
\end{multline}
\end{subequations}
 \begin{figure}
 	\centering
 	\includegraphics[width=0.7\linewidth]{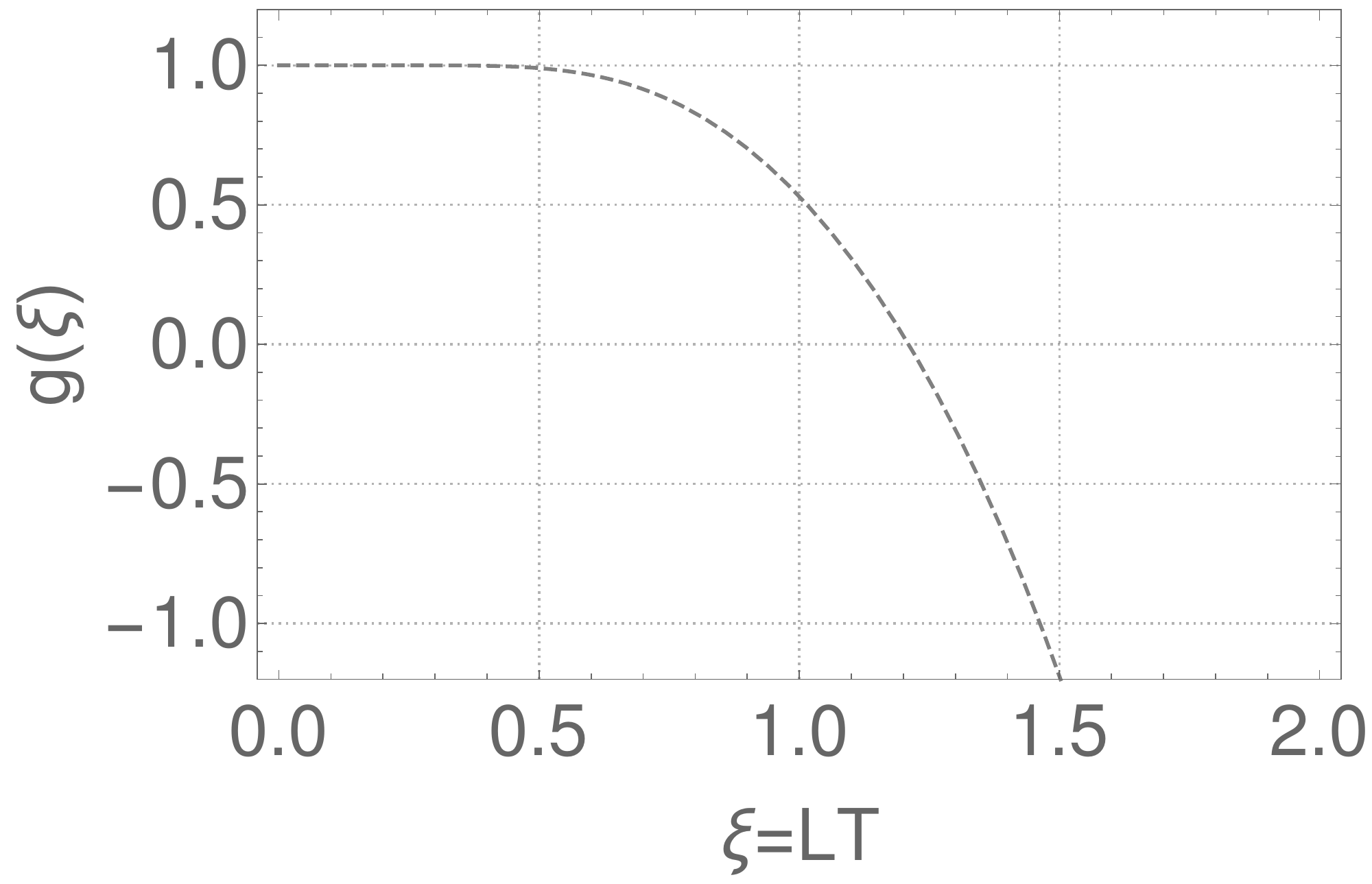}
 	\caption{Behavior of the function $g(\xi)$, for fermions, using antiperiodic boundary conditions. Curve plotted using 500 terms of the double sum. Curve plotted using 500 terms of the double sum. The root of the function, $g(\xi^*)=0$, is around $\xi \sim 1.21 \hbar c/k_B = 239/k_B \text{fm} \cdot \text{MeV}$  (back into SI units).}
 	\label{fig:fermionicltperiodic}
 \end{figure}
\begin{figure}
	\centering
	\includegraphics[width=0.7\linewidth]{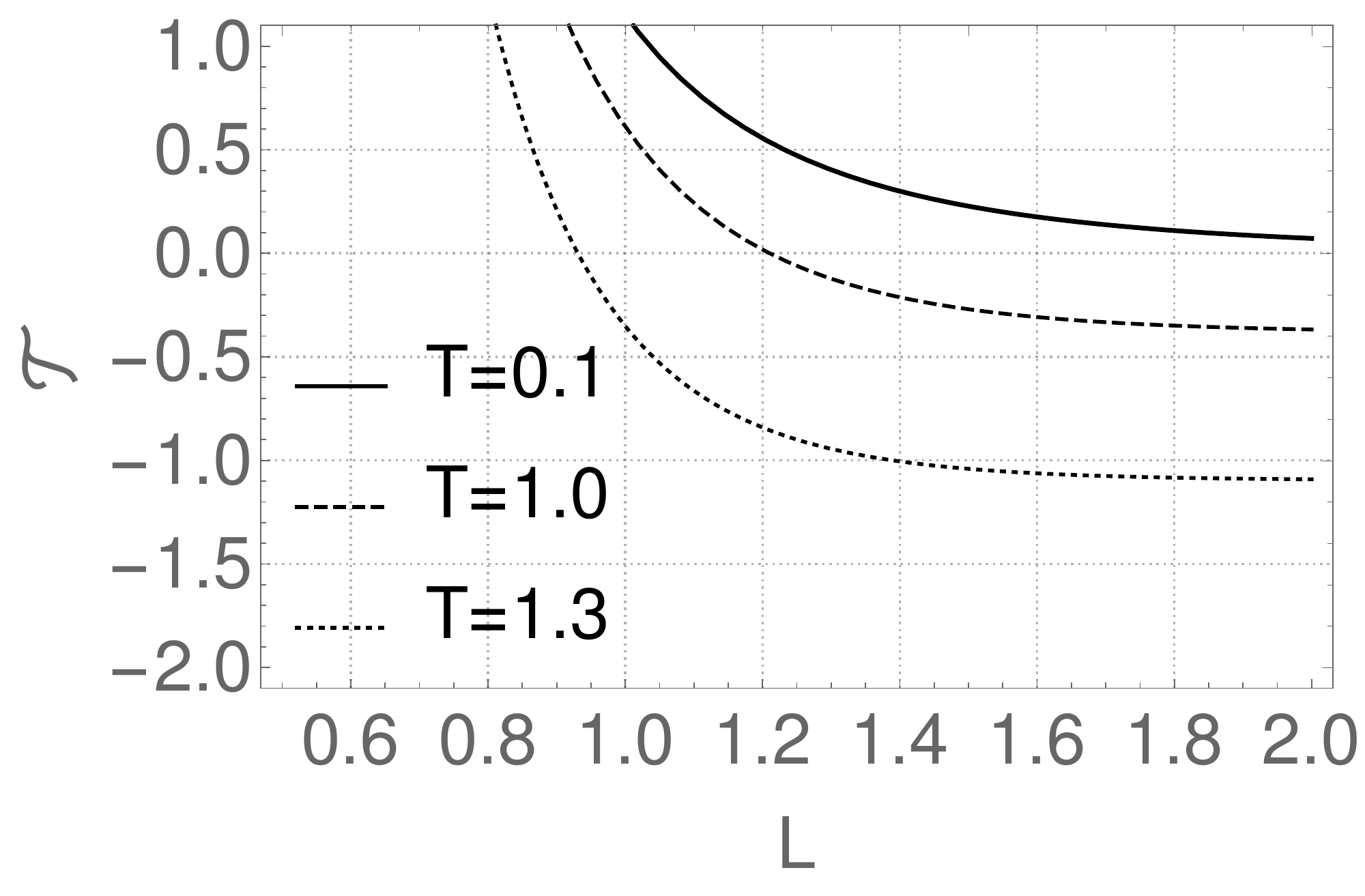}
	\caption{Strength of the force with respect to the size of the cavity for different heat bath temperatures (for fermions, using antiperiodic boundary conditions in space). With the increase of the temperature we get a length at which the force becames attractive}
	\label{fig:fermioniccurveltperiodic}
\end{figure}
The Casimir force for fermions in a cavity at a thermal bath is attractive, as expected, for periodic boundary conditions in space. We must remark that the thermal contribution does not produce any compensation as we observed for the scalar scenario. That is, regardless of the temperature the force is expected to stay attractive. However, when we assume antiperiodic boundary conditions in space the Casimir force produced by the quantum vacuum starts as repulsive (just as we observe for scalar fields) and the increase in temperature helps to turn it back into an attractive force. This flip in the sign of the force is due to the change in the sign of the function $g(\xi)$ as we can see in Fig.~\ref{fig:fermionicltperiodic}. The higher the temperature $T$, the smaller the cavity size $L$ one requires to flip the sign of the force, see Fig.~\ref{fig:fermioniccurveltperiodic}. 

\begin{figure}
	\centering
	\includegraphics[width=0.7\linewidth]{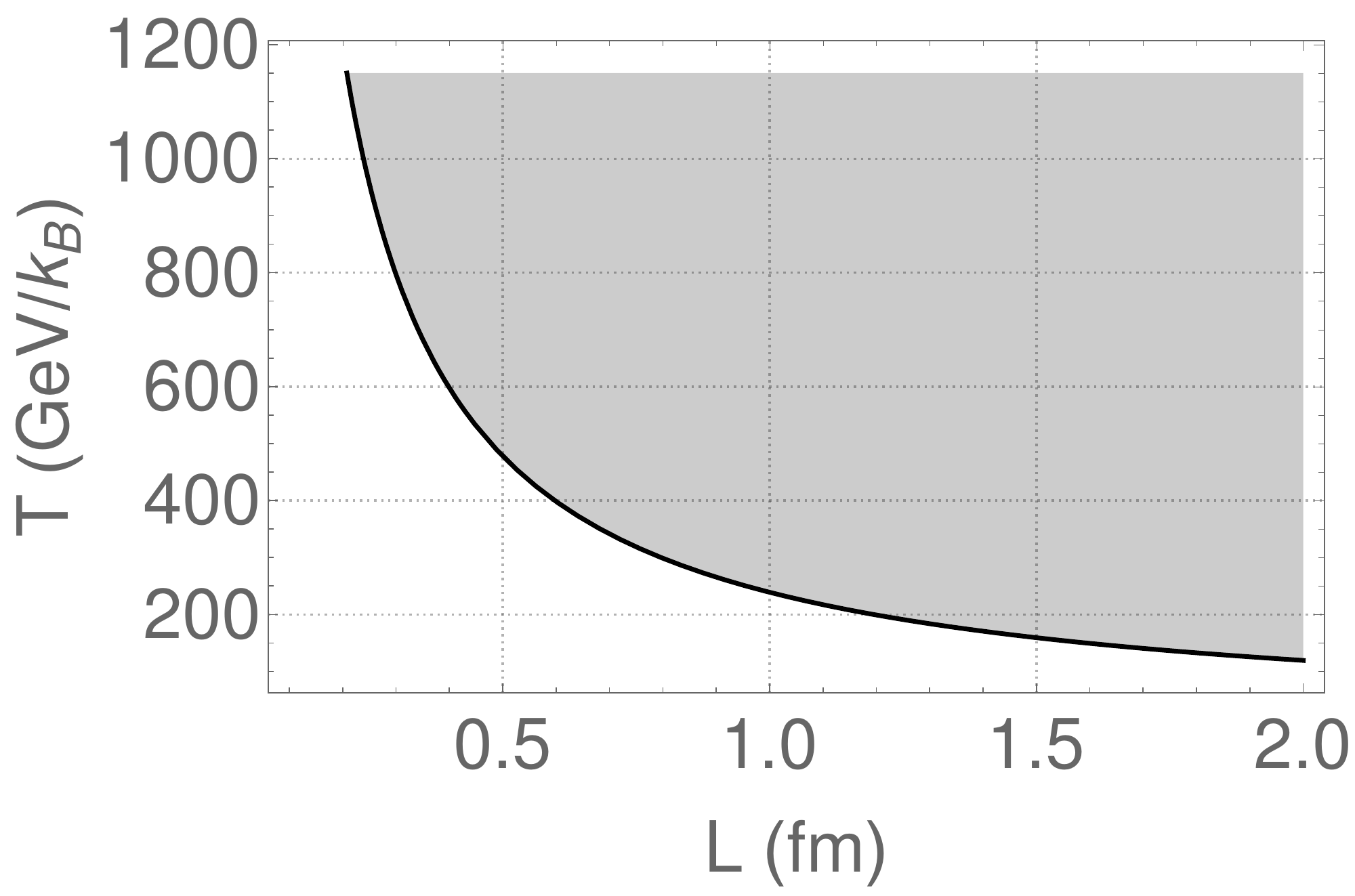}
	\caption{Diagram showing the change of the sign of the Casimir force for fermionic fields under antiperiodic boundary conditions. The filled region represents attractive Casimir forces.}
	\label{fig:fermionphasediagram}
\end{figure}

Notice that the characteristic point the the Casimir force vanishes occurs at $L \simeq 1.21/T$. Or, back into SI units, $LT \simeq 1.21 \hbar c/k_B = 2.77 \times 10^{12} \text{fm} \cdot \text{K} = 239/k_B \text{fm} \cdot \text{MeV}$. Meaning that to a cavity around the typical size of a meson (1 fm) the temperature at which the Casimir force vanishes is around $2.77\times10^{12}$K (or $239$MeV). We illustrate the `phase diagram' for fermionic field under antiperiodic boundary condition in Fig.~\ref{fig:fermionphasediagram}

\section{Final remarks}
\label{finalremark} 
Our results tell us that:
	\begin{enumerate}
		\item fermions with periodic boundary conditions (b.c.), are always subject to attractive quantum forces;
		\item bosons with antiperiodic b.c. always are subject to repulsive quantum forces;
		\item bosons with period b.c. are subject to attractive forces  at small distances or low temperatures and repulsive at large distances or high temperatures. This means that there is an `equilibrium configuration point' at $LT\simeq (300/k_B) \text{MeV}\cdot \text{fm} \approx 3.49 \text{mm}\cdot \text{K}$ where the force  vanishes; however, this point is unstable: Any small deviation makes the system collapse or expand;
		\item fermions with antiperiodic b.c. feel  an attractive force at large distances or low temperatures and a repulsive one at small distances or high temperatures. This means that we have again an `equilibrium point' (at $LT\simeq (239/k_B) \text{MeV}\cdot \text{fm} \approx 2.77 \text{mm}\cdot \text{K}$), but this time this point is stable.
	\end{enumerate}

In the present work, we propose an alternative way to investigate and compute quantum forces in the context of field theories in nontrivial topologies. In particular, employing the generalized Matsubara formalism, we recover the so-called Casimir pressure considering a massive scalar field in a compact space in thermal equilibrium.

The usual attractive response of quantum and thermal fluctuations is obtained and our results are in accordance with
those found in the literature. One may notice that all thermal contributions
to the Casimir pressure, given by $\mathcal{T}_{n_{2}}^{\text{scalar}}$ and 
$\mathcal{T}_{n_{1}n_{2}}^{\text{scalar}}$, vanish in the
zero-temperature ($\beta \rightarrow \infty $) limit, remaining the pure
dependence on the distance $L$ between plates, which has the well-known 
$L^{-4} $ dependence in the small-$L$ limit for a four-dimensional space.
Also, the bulk limit $L\rightarrow \infty $ reduces all expressions in $D=4$
to the Stefan--Boltzmann law $\beta ^{-4}$.

A rather peculiar aspect of the generalized Matsubara formalism is related
to the renormalization of the expressions. Usually, in the Casimir context,
the divergent terms are taken care of by subtraction of the bulk integral,
without compactifications (see \cite{Khanna-et-al-Phys-Rep-2014}). Here,
there is no need to do so, as was also remarked by Elizalde and Romeo \cite
{Elizalde-Romeo-JMP-1989, Elizalde-Romeo-JMP-1990}. It is sufficient to obtain correct physical
expressions to renormalize them by subtracting the divergent term of the
expansion of the Epstein--Hurwitz zeta functions $Z_{d}^{c^{2}}$, as it does
not depend on the physical parameters $L$ or $\beta $. 

We also remark that the expression we obtain from the field theory in toroidal spaces formalism, which conveys periodic boundary conditions in the compactified dimensions, leads to corresponding ones for the Dirichlet conditions, by substituting $L=2a$. The $D=4$, the small-$L$ limit of the Casimir pressure in the nonthermal case, given by Eq.~(\ref{Casimireffect-compactifiedspace}) becomes $\mathcal{T}_{33}=-\pi ^{2}/480a^{4}$ in the Dirichlet case for a quantum
scalar field. For an electromagnetic field, we have then twice that value, $\mathcal{T}_{33}=-\pi ^{2}/240a^{4}$, due to its two degrees of freedom. These are compatible with the original Casimir results.

In summary, our results are in accordance with those found in the literature, obtained from other techniques, and also illustrate the relevance and interdisciplinarity of the generalized Matsubara procedures as a theoretical platform to deal with quantum phenomena in nontrivial topologies.

\section{Acknowledgments}
\label{ack} 
This work was partially supported by the Brazilian agencies CNPq and FAPERJ. 

\end{document}